\documentclass[12pt,onecolumn]{article}
\usepackage[authoryear,comma]{natbib}
\usepackage{amsmath}
\usepackage{amssymb}
\usepackage{graphicx}
\usepackage{euscript}
\newcommand{\beq}{\begin{equation}}
\newcommand{\eeq}{\end{equation}}

\title{Time variability of viscosity parameter in differentially
rotating discs}
\author{S. R. Rajesh$^{1}$ \& Nishant K. Singh$^{2,3,\footnote{Corresponding author. Email: nishant@nordita.org\;; Fax: +46 8 5537 8404}}$ \\[3ex]
$^{1}$S. D. College, Allappuzha, Kerala 688 003, India\\[2ex]
$^{2}$Inter--University Centre for Astronomy and Astrophysics, \\Post Bag 4,
Ganeshkhind, Pune 411 007, India\\[2ex]
$^{3}$NORDITA, KTH Royal Institute of Technology and Stockholm University,\\
Roslagstullsbacken 23, SE-10691 Stockholm, Sweden}
\date{}

\begin{document}

\maketitle

\begin{abstract}
We propose a mechanism to produce fluctuations in the viscosity parameter ($\alpha$)
in differetially rotating discs. We carried out a nonlinear analysis of a general
accretion flow, where any perturbation on the background $\alpha$ was treated as a
passive/slave variable in the sense of dynamical system theory. We demonstrate a
complete physical picture of growth, saturation and final degradation of the perturbation
as a result of the nonlinear nature of coupled system of equations.
The strong dependence of this fluctuation on the radial location in the accretion disc
and the base angular momentum distribution is demonstrated. The growth of fluctuations is 
shown to have a time scale comparable to the radial drift time and hence the physical
significance is discussed. The fluctuation is found to be a power law in time in
the growing phase and we briefly discuss its statistical significance.
\end{abstract}


\section{Introduction}

The presence of accretion discs around compact objects like the neutron star and the black hole
in both galactic and extragalactic X-ray sources is now a well established phenomenon.
Some radio objects such as active galactic nuclei have accretion discs around supermassive black holes. Apart from several details such as environment, size, 
strength of magnetic field, cooling mechanism etc., all the global models of accretion systems share
a common hydrodynamic structure. The central idea being, the 
turbulent shear stress causes dissipation of angular momentum and energy of the rotating fluid particles such that accretion can take 
place. The origin of turbulence and hence turbulent viscosity was an issue for the founders of the field of `accretion powered 
astrophysical systems' and still remains to be a major issue. The closure model proposed by \cite{SS73} remains the only working model 
for turbulent shear stress in astrophysical accretion discs. In this model, the physics behind the turbulent shear stress 
is parametrised by a dimensionless number $\alpha$. Thus the $\alpha$ viscosity continues to be the central idea in any 
model for hydrodynamic transport in accretion systems. 

The spirit of the $\alpha$ viscosity is as follows: any eddy velocity which is greater than the local sound speed will dissipate 
quickly and cannot be the cause of eddy viscosity. Hence the turbulent stress must be less than the local isotropic pressure. 
Thus the shearing stress is taken to be proportional to the local isotropic pressure where the proportionality factor is 
called $\alpha$, where $0 < \alpha <1$. When $\alpha \sim 1$ the flow is called a high viscosity flow, whereas when 
$\alpha \lesssim 10^{-2}$, the flow is called a low viscosity flow. With this model, the spectrum of cool Keplerian discs could be 
explained \citep{PR72, NT73, SS73}. The idea of a sub-Keplerian disc was proposed to explain the nonthermal tail of the spectra 
from X-ray sources \citep{LT80, PB81, MP82}. In the case of a sub-Keplerian accretion disc the turbulent energy 
dissipated locally, is partly advected radially and partly emitted as radiation via nonthermal processes. In this model also, 
$\alpha$ closure remained unchanged although additional ram pressure was added to 
the total pressure \citep{CT95, MC05, RM10}. Since $\alpha$ is the ratio of two flow variables, namely, the turbulent
shear stress and isotropic pressure, $\alpha$ should also be considered as a flow (continuum) variable. 
Since there is no known equation for the evolution of $\alpha$, it is treated as a disc parameter, and its value is fixed globally.

Apart from the fundamental problem to explain the origin of turbulent viscosity, there are other phenomena which await 
complete understanding, such as rapid X-ray variabilities in black hole accretion discs, aperiodic X-ray fluctuations and
quasi periodic oscillations (QPOs) in accretions discs. Global mode oscillations and waves are invoked to
explain some of these phenomena \citep{Muk09}. As the $\alpha$ viscosity is the source of 
energy dissipation in accretion discs, it is logical to attribute some of the time variabilities of the spectra to the temporal variation of 
$\alpha$. For example, in order to explain the $1/f$ (flicker) noise in X-ray sources, \cite{Lyu97} considered the local temporal 
fluctuations of $\alpha$ at outer radii. This would cause a change in mass accretion rate
at inner radii where most of the 
X-rays are emitted. In order to have such an effect, a time varying component of $\alpha$ was assumed. The fluctuation was assumed 
to grow enough in a local accretion time scale. Thus the fluctuations in $\alpha$ resulted in a variable mass accretion rate, which would
lead to variations in X-ray luminosity. Although the ultimate aim of our work in the present paper is similar to that of
\cite{Lyu97}, i.e., to study a mechanism to produce variabilities in the observed luminosities (say, X-ray) from accretion discs,
our approach is somewhat different, and may be stated as follows:
considering a steady state accretion in an annular region of an accretion disc with self-similar base flow profiles, we wish to
study how the background $\alpha$ changes in response to any perturbation on the radial velocity field ? Such perturbations
of the radial velocity field (i.e., the mass accretion rate) may, in general, be of internal or external origin. These kind
of studies have direct implications on the observed variabilities in X-ray luminosities from accretions discs.

A stable accretion system tries to maintain the steady mass flow across all radii. Any cause, internal or external which 
disturbs this steady state, will be quickly nullified by viscous dissipation. In \textsection~(2.1) we model the steady state flow 
variables in a local annular region as a power law in radial coordinate. The global flow domain can be thought of as a collection of such 
annular regions. In \textsection~(2.2) the evolution equations for perturbations in the mean density and the radial velocity, causing
perturbation in $\alpha$ are discussed. We use the standard $\alpha$ model for viscous stress.
In \textsection~(2.3) we reduce the perturbation equations to a set of nonlinear dynamical systems of equations,
by specializing to the case when the Lagrangian derivatives (defined with respect to the radial velocity field)
of the perturbations in the flow variables vanish. In \textsection~(3) we demonstrate that the growth of the
viscosity parameter is always followed by saturation and degradation, and the fluctuation asymptotically goes to zero. The behaviour of the
fluctuation is strongly dependent on the radial location, base angular momentum distribution and the mean viscosity of the flow.
We demonstrate that the growth of the fluctuation in viscosity parameter always scales as the local accretion time, and that
it shows a power law growth phase in time, in the astrophysically relevant time scale. We conclude in \textsection~(4).

\section{Model Equation Describing The System And The Solution Procedure}

Let us consider a cylindrical coordinate system, with spacetime coordinates denoted by $(r, \phi, z, t)$, whose origin
is at the centre of the compact object. The angular velocity vector, ${\bf \Omega}$, is pointed along $z$ (vertical) direction, and
the midplane of the accretion disc is at $z=0$. We begin by considering a vertically integrated, axisymmetric, steady-state accretion flow,
in which, we focus on an annular region of the accretion disc. We consider axisymmetric perturbations on base radial
velocity field and mean density in this annular region to study the response of such perturbations on the evolution of the
viscosity parameter. Thus all the base flow variables in this study are functions only of the radial coordinate ($r$), whereas the perturbations
depend on both, the radial coordinate ($r$) and time ($t$).

\subsection{Base flow}

For a general accretion flow, we consider a small annular region specified by the mean velocity field,
where $V$ and $\Omega$ are the magnitudes of the radial and angular velocity fields, respectively.
Let us specify the unperturbed axisymmetric, steady-state accretion (base) flow where the radial velocity and the
angular velocity are power laws in radial coordinate, i.e., $V = V_{0} \, r^{-j}$ and 
$\Omega = \Omega_{0} \, r^{-q}$. The explicit radial dependence of other fluid variables in an unperturbed state
can be obtained by solving the conservation equations for mass, radial momentum and angular momentum, given as:

\beq
\frac{1}{r}\frac{\partial}{\partial r}\,(r\Sigma V) \;=\; 0
\label{UP-Cont}
\eeq

\beq
\Sigma \left(V\frac{\partial V}{\partial r}\,-\,r \Omega^2 \right)\;=\;
\frac{k \Sigma}{r^2}\,-\,\frac{\partial P}{\partial r}
\label{UP-RadMom}
\eeq

\beq
\frac{\Sigma V}{r^2}\frac{\partial}{\partial r}(r^2 \Omega) \;=\; -\frac{1}{r^3}
\frac{\partial}{\partial r}(r^3 W^{\phi r})
\label{UP-AziMom}
\eeq

\noindent
where $\Sigma$ and $P$ are vertically integrated density and pressure, respectively. The quantity $k = -G M$, where $G$ is the universal 
gravitational constant and $M$ is the mass of the central object. We solve the above set of equations along with the equation of 
state, $P \, = \, \Sigma  \, T$, where $T$ is the effective temperature of the flow. We impose the boundary condition that all the physical quantities
go to zero as $r \, \rightarrow \infty$. For the turbulent stress, we use the $\alpha$ 
viscosity model, i.e., $W^{\phi r} = \alpha \, P / r$, where $\alpha$ is the Shakura-Sunyaev viscosity parameter. 
We can write the solution to the above set of equations as, 

\beq
V(r) \,=\, V_{0} \, r^{-j}\;;\qquad \Omega(r) \,=\, \Omega _{0} \, r^{-q}\;;\qquad \Sigma(r) \,=\, \Sigma_{0} \, r^{j-1}
\label{BFsoln}
\eeq

\beq
T(r) \,=\, \left(\frac{k}{j-2}\right) r^{-1} \,-\, \left(\frac{j V_{0}^2}{j+1}\right) r^{-2j} \,+\,
\left(\frac{\Omega_{0}^2}{j-2q+1}\right) r^{2(1-q)}
\label{BF-Tsoln}
\eeq

\noindent
where $V_0 < 0$, as the radial flow is directed towards the central object, and $\Sigma_{0} = \dot{M}/(4 \pi V_0)$,
where $\dot{M}$ is a negative quantity called the mass accretion rate. Since both, the radial velocity and density, decrease with increasing
values of $r$, we get from Eq.~(\ref{BFsoln}) that $0 < j < 1$. The value of $q$ indicates the angular momentum distribution of the
base flow; $q \, = \, 1$, $q \, = \, 3/2$ and $q \, = \, 2$ describe, respectively, the flat rotation, Keplerian rotation
and constant angular momentum disc profiles. From the angular momentum balance equation, we get 

\beq
\alpha(r) T(r) \,=\, -V_{0} \Omega_{0} r^{1-q-j} \,+\, \frac{c}{\Sigma_0} r^{-j-1}
\label{BF-alpTsoln} 
\eeq

\noindent
As $\alpha T$ is a physical quantity, it approaches zero as $r \, \rightarrow \infty$, according to the boundary condition
that we have chosen. In Eq.~(\ref{BF-alpTsoln}), since the term containing the integration constant $c$ goes to zero 
as $r \, \rightarrow \infty$, $c$ is nonzero in general. The
physics behind $c$ comes from the actual physical mechanism which produces the $\alpha$ viscosity. The origin of $\alpha$ 
is beyond the scope of the present analysis, therefore, we can only choose $\alpha$ at a particular radius which automatically
fixes the value of $c$ from Eq.~(\ref{BF-alpTsoln}). 

\subsection{Perturbation}

Any closure model in hydrodynamics is based on the assumption that the stress tensor can be written as a functional of 
space and time through the mean flow variables, such as mean density, velocity etc. In the case of an accretion disc it is this unknown dependence which is characterized by $\alpha$. Thus 
the perturbation in mean hydrodynamic quantities should cause an effective change in $\alpha$ or vice versa due
to their dynamical coupling. The evolution of $\alpha$ cannot be traced rigorously since we do not yet have any equation
describing $\alpha$. In the following perturbation analysis, we perturb the
mean density and radial velocity fields, which cause an effective change in the mass accretion rate, thus effectively changing the
angular momentum flux. In the present work, we explore the scenario in which the stress tensor fluctuates instantaneously, whereas
the angular momentum distribution remains unchanged. Assuming that $\sigma$, $v$ and $\beta$ denote perturbations on the base quantities
$\Sigma$, $V$ and $\alpha$ respectively, such that, $\Sigma \, \rightarrow \, \Sigma \, + \, \sigma$, $V \, \rightarrow \, V \, + \, v$ and 
$\alpha \, \rightarrow \, \alpha \, + \, \beta$, we can write the following equations for the perturbations:

\beq
\frac{\partial \sigma}{\partial t}\,+\,\frac{1}{r}\frac{\partial}{\partial r}
\left[r \Sigma v\,+\,r \sigma V\,+\,r \sigma v\right] \;=\; 0
\label{P-Cont1}
\eeq

\beq
(\Sigma + \sigma)\frac{\partial v}{\partial t}\,+\,\sigma V \frac{\partial V}{\partial r}
\,+\,(\Sigma + \sigma)\left[ V \frac{\partial v}{\partial r}\,+\,
v \frac{\partial V}{\partial r}\,+\,v \frac{\partial v}{\partial r}\right]\,-\,
\sigma r \Omega^2 \;=\; \frac{k \sigma}{r^2}\,-\,\frac{\partial}{\partial r}(\sigma T)
\label{P-RadMom1}
\eeq

\beq
\left[ \frac{\sigma V}{r^2}\,+\,\frac{(\Sigma + \sigma) v}{r^2} \right]
\frac{\partial}{\partial r}(r^2 \Omega) \;=\; -\frac{1}{r^3}\frac{\partial}{\partial r}
\left[ r^2 (\alpha \sigma \,+\, \beta \Sigma \,+\, \beta \sigma) T \right]
\label{P-AziMom1}
\eeq

\noindent
We note that the perturbations are functions of $r$ and $t$, as discussed above, i.e., $\sigma=\sigma(r, t)$, $v=v(r, t)$ and
$\beta=\beta(r, t)$.

\subsection{Dynamical model}

Any general accretion disc is expected to possess all sorts of perturbations, the exact
form of which could be completely unknown and hence arbitrary.
We consider the perturbation such that the observer, moving along with a fluid particle radially, sees no change in the 
perturbation, i.e., the perturbation is frozen to the radially advecting fluid particle. Thus the Lagrangian or
advective time derivative of the perturbation is zero. We define the advective time derivative with respect to base radial flow, which
is denoted as,

\beq
\frac{d}{dt} \,\equiv\, \frac{\partial}{\partial t} \, + \, V\frac{\partial}{\partial r}
\label{AdvDer}
\eeq

\noindent
Let $r_0$ be an arbitrary length scale. We scale the radial coordinate by $r_0$ and denote
the scaled radial coordinate as $\widetilde{r} = r/r_0$. We scale the other variables
as,

\beq
\widetilde{V}\,=\,\frac{V}{V_0 r_0^{-j}}\,=\,\widetilde{r}^{\,-j}\;;\quad
\widetilde{\Sigma}\,=\,\frac{\Sigma}{\Sigma_0 r_0^{j-1}}\,=\,\widetilde{r}^{\,j-1}\;;\quad
\widetilde{\Omega}\,=\,\frac{\Omega}{\Omega_0 r_0^{-q}}\,=\,\widetilde{r}^{\,-q}
\eeq

\noindent
We scale the time and the temperature as,

\beq
\widetilde{t}\,=\,\frac{t}{\left(\Omega_0 r_0^{-q}\right)^{-1}}\,=\,
(\Omega_0 r_0^{-q})\,t\;;\qquad \widetilde{T}\,=\,\frac{T}{\left(\Omega_0 r_0^{1-q}\right)^2}
\label{t-scale}
\eeq

\noindent
Where $r_0 \Omega(r_0) = r_0 \Omega_0 r_0^{-q} = \Omega_0 r_0^{1-q}$ is the azimuthal velocity
at distance $r_0$. We scale the perturbed quantities $\sigma$ and $v$ as following:

\beq
\widetilde{\sigma}\,=\,\frac{\sigma}{\Sigma_0 r_0^{j-1}}\;;\qquad
\widetilde{v}\,=\,\frac{v}{V_0 r_0^{-j}}
\eeq

\noindent
Thus we scale Eq.~(\ref{BF-Tsoln}) and write the following expression for temperature in scaled units:

\beq
\widetilde{T}(\widetilde{r}\,)\;=\;\left(\frac{\epsilon_1}{j-2}\right) \widetilde{r}^{\,-1}\,+\,
\left(\frac{1}{j-2 q+1}\right)\widetilde{r}^{\,2(1-q)}\,-\,\left(\frac{\epsilon_2\, j}{j+1}\right)\widetilde{r}^{\,-2j}
\label{BF-Tscaled}
\eeq

\noindent
where,

\beq
\epsilon_1\;=\;\frac{k\, r_0^{-1}}{\left(\Omega_0 r_0^{1-q}\right)^2}\;;\qquad
\epsilon_2\;=\;\frac{\left(V_0 r_0^{-j}\right)^2}{\left(\Omega_0 r_0^{1-q}\right)^2}
\label{eps12}
\eeq

\noindent
We note that $\epsilon _{1}$ is the ratio of the local gravitational energy to the angular kinetic energy which
is a negative number; $|\epsilon _1| \simeq 1$ for a Keplerian accretion disc and $|\epsilon _1| > 1$ for a sub-Keplerian
disc. From Eq.~(\ref{eps12}), we see that $\epsilon _2$ is the ratio of the local radial kinetic energy to the angular kinetic energy;
$\epsilon _2 << 1$ for a Keplerian accretion disc and $\epsilon _2 > 1$ for inner sub-Keplerian disc very close to the central object.
Without loss of generality we can choose $\widetilde{r}\, = \, 1$.
The information of radial location in the disc is implicitly contained in the parameters $j$, $q$, $\epsilon_1$ and $\epsilon_2$,
and thus one can sample different radial locations by suitably varying these parameters. Using Eq.~(\ref{AdvDer}) and the scaling relations in Eqns.~(\ref{P-Cont1})-(\ref{P-AziMom1}) we
find that the set of dynamical equations for perturbations then reduces to

\begin{eqnarray}
\label{sig}
\frac{\partial \widetilde{\sigma}}{\partial \widetilde{t}} &=&
F_1(\widetilde{\sigma}, \widetilde{v};\, j, q, \epsilon_1, \epsilon_2)\\[2ex]
\label{v}
\frac{\partial \widetilde{v}}{\partial \widetilde{t}} &=&
F_2(\widetilde{\sigma}, \widetilde{v};\, j, q, \epsilon_1, \epsilon_2)\\[2ex]
\label{bet}
\frac{\partial \beta}{\partial \widetilde{t}} &=& 
F_3(\widetilde{\sigma}, \widetilde{v}, \beta;\, j, q, \epsilon_1, \epsilon_2)
\end{eqnarray}

\noindent
Details of the functions $F_1 \, , \, F_2 \, , \, F_3$ are given in the Appendix B. 

\section{Global behaviour of dynamical system}

\begin{figure}
\centering
\includegraphics[width=0.90\columnwidth]{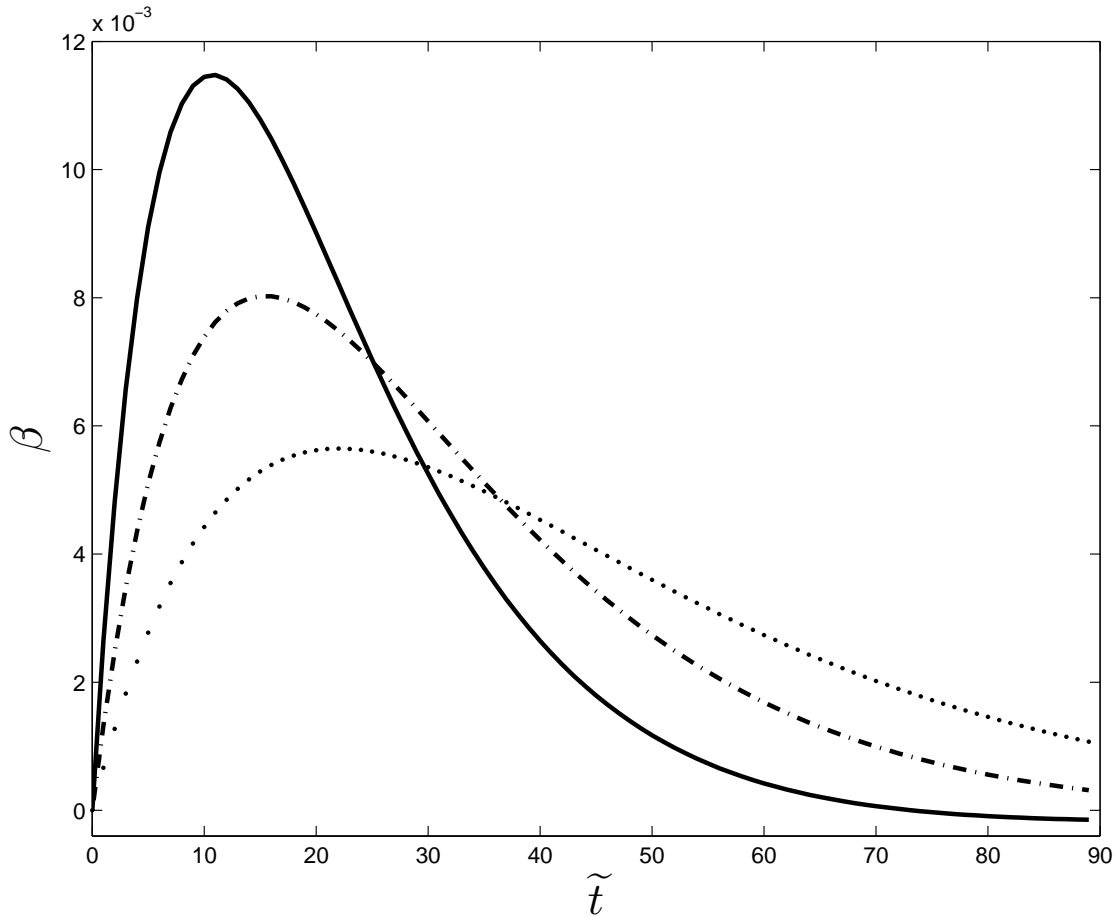}
\caption{The temporal variation of $\beta$ is plotted corresponding to the Keplerian disc (i.e., $q=1.5$) for $\alpha \simeq 10^{-2}$,
$\epsilon_1 = -1.1$ and $j=0.9$. The solid bold line corresponds to 
$\epsilon_2 = 0.01$, the dashed-dotted line corresponds to $\epsilon_2 = 0.005$ and the dotted line is for $\epsilon_2 = 0.0025$. The horizontal axis denotes the time measured
in units of the rotational time-scale (see Eq.~(\ref{t-scale})).}
\label{beta-t-Kep}
\end{figure}

Our aim is to solve the system of coupled nonlinear dynamical equations,
(\ref{sig})--(\ref{bet}), as an initial value problem. The density ($\sigma$)
and radial velocity ($v$) perturbations are nonlinearly coupled to each other,
whereas the dynamical evolution of $\beta$ is governed by its nonlinear coupling
to $\sigma$ and $v$, as may be seen from Eq.~(\ref{bet}). The quantity $\beta$ may be
thought of as a passive/slave mode in the sense of dynamical system theory \citep{Man90}.
We wish to study a mechanism which can produce temporal variations in the viscosity
parameter, $\alpha$. Therefore $\beta$, signifying any departure from a chosen
base value of $\alpha$, is universally assumed to be zero at an initial time. The
initial density perturbation is also chosen to be zero throughout.
In this setup, an initial fluctuation of the radial velocity is assumed and we
analyze the evolution of $\sigma$, $v$ and $\beta$ as a function of time, for
different values of free parameters, $j$, $q$, $\epsilon_1$ and $\epsilon_2$. In Fig.~(\ref{beta-t-Kep})
a typical plot of $\beta$ as a function of time is shown for the Keplerian
angular velocity profile for different values of $\epsilon_2$, keeping everything else same. 
When the system is pumped with a random external velocity fluctuation the 
density and radial velocity fields change. Since $\beta$ is dynamically coupled to the
density and velocity fluctuations, it gets exited and shows growth in the present example. We can see from Fig.~(\ref{beta-t-Kep})
that $\beta$ grows with time, reaches a saturation and finally degrades.
The major difficulty with standard linear perturbation analysis is that they could
either show growth or decay. A complete physical picture of growth, saturation and final
degradation of the perturbation is the result of the nonlinear nature of the system of equations, which is a non-trivial result.

When the system is pumped with a random initial radial velocity fluctuation, the density and radial velocity perturbation 
evolve according to the rule governed by Eqns.~(\ref{sig})--(\ref{v}). We observe that for the physically acceptable range of 
free parameters  $j$, $q$, $\epsilon_1$ and $\epsilon_2$ (the allowed range of  $j$, $q$, $\epsilon_1$ and $\epsilon_2$ will be discussed 
in the next subsection) all initial values form orbits attracted towards origin ($\sigma \, = \, 0$ , $v \, = \, 0$) in the $\sigma \, - \, v$ 
plane. In other words, for the physically acceptable range of free parameters, we find that $\sigma \, = \, 0$ and $v \, = \, 0$ is 
a universal attractor. This observation along with Eq.~(\ref{bet}) means that asymptotically $\beta$ is always zero.

The physical reason behind the time variability of $\beta$ may be explained as follows: as the viscous stress dissipates the angular momentum 
the fluid particle gains more radial drift velocity and moves to the inner orbit. The radial drift velocity towards the central object 
is larger for a high viscosity flow (larger $\alpha$) than a low viscosity flow (smaller $\alpha$); see, for example, \cite{RM10}.
When a positive radial velocity component is added to the fluid particle, the particle moves with less radial drift velocity towards the
central object. The tendency of the fluid particle to remain at a larger orbit is thus overcome by increasing viscous
dissipation and hence $\beta$ shows increasing 
tendency. Finally when the density/radial velocity perturbations damp, the $\beta$ asymptotically becomes zero. The reverse is true 
when a negative initial radial velocity fluctuation is added to the system. The spatial dependence of the evolution of the perturbation is 
contained in the parameters of the system such as $q,j,\epsilon_1,\epsilon_2$. In Fig.~(\ref{beta-t-Kep}), we demonstrate the
evolution of $\beta$ as a response to a positive radial velocity perturbation in a Keplerian disc for different values of $\epsilon_2$, while
keeping rest of the parameters the same. We note that different values of $\epsilon_2$, while keeping other parameters the same, represent
different radial distances in the accretion disc, with larger values corresponding to the smaller orbits. Therefore the bold, the dashed-dotted
and the dotted curves in Fig.~(\ref{beta-t-Kep}) correspond to the innermost, the intermediate and the outermost radial location from the
central object. Thus we find that, for the same initial trigger, the stronger fluctuation in viscosity is seen for relatively inner orbit. The 
reason for this nature could be understood as follows: the fluid particle stays in the inner orbit for less time as compared to the outer 
orbit. So the fluid particle in the inner orbit gets relatively less time to neutralise the external energy input. To dissipate this 
energy the viscosity has to increase much larger in relatively smaller times. 

\subsection{Solution of dynamical system}

We have a whole range of parameters in this problem, whose physical meanings could be briefly stated as follows:
(i) $q$ determines the distribution of base angular momentum; (ii) $\epsilon_1$, which is the ratio of gravitational energy to 
the angular kinetic energy of the fluid particle at a particular radius, signifies the competition between gravity and angular momentum,
with value unity corresponding to strict balance of gravity and angular momentum as in the case of the Keplerian motion;
(iii) $\epsilon_2$, which is the ratio of radial kinetic energy to the angular kinetic energy, represents the strength of advection,
and is expected to be larger in the inner parts of the accretions discs where the flow
could be advection dominated depending 
on the efficiency of the cooling mechanism and strength of turbulent viscosity; and
(iv) $j$ represents the base distribution of density and radial velocity fields.
We can find the Bernoulli number ($B$) of the flow by expressing the radial velocity equation in integral form. We can also define the 
Mach number ($M$) as the ratio of radial speed to the local sound speed. In the scaled units at $\widetilde{r} \, = \, 1$, $B$ and 
$M$ have the following form:

\begin{eqnarray}
\nonumber
B &=& \frac{\epsilon_2}{2} + \frac{1}{2\left(q \, - \, 1\right)} + \epsilon_1 + \widetilde{T} \, + \,
\left(j\,-\,1 \right) \left[ \frac{\epsilon_1}{\left(2\,-\,j\right)} \, +\,  
\frac{1}{\left(j\,-\,2q\,+\,1\right)\left(2\,-\,2q\right)} \,+\,\frac{\epsilon_2}{2 \left(j\,+\,1\right)} \right]\\[2ex] \nonumber
M &=& \sqrt{ \frac{\epsilon_2}{\widetilde{T}}}
\label{BM}
\end{eqnarray}

\subsubsection{The parameter search}

As our system of equations (Eqns.~(\ref{sig})-(\ref{bet})) depends on many parameters discussed above, we need to carefully
choose the values of these parameters in order to get physically acceptable solutions. Many of these parameters are restricted and should
be chosen such that they satisfy the conditions, $T>0$ and $B<0$. We note from Eq.~(\ref{BFsoln}) that $0<j<1$, and also restrict ourselves to cases
when $1.5 \leqq q < 2$. We are not considering the effect of shocks which may be significant at much of the inner part of the 
accretion disc. Hence throughout the analysis we consider subsonic flow, i.e.; $M \, < \, 1$. 
We search for some suitable values of these parameters in the following two regimes:

\begin{enumerate}
\item [(i)] We consider outer regions of an accretion disc such that the value of $q$ is close to the Keplerian value ($q \simeq 1.5$).
Thus we explore the cases when $1.5 \leqq q < 1.7$. For such a disc, we consider both, the low viscosity ($\alpha \lesssim 10^{-2}$) 
and high viscosity ($\alpha \sim 1$) flows. Having chosen these parameters, we now study our set of dynamical equations as a function of remaining
parameters, where we always ensure that our choices are consistent with conditions, $T>0$, $B<0$ and $M \, < \, 1$.
\item [(ii)] Next, we focus on regions of the accretions disc which have dissipated much of their angular momentum, like, for example,
inner parts of the disc, where the value of $q$ would be closer to the case of constant angular momentum disc, for which $q=2$. Thus we explore
the cases when $1.7\leqq q<2$. Again, we study low viscosity and high viscosity discs in a self-consistent manner as discussed above.
\end{enumerate}

\noindent
In the standard Keplerian disc proposed by \cite{SS73} there is an exact balance between the gravitational force and the centrifugal 
reaction experienced by the rotating fluid particle. Therefore for a flow with $q$ more close to the Keplerian value the magnitude 
of $\epsilon_{1}$ should be more close to unity. In this case the radial drift velocity is too small compared to the corresponding 
rotational velocity. So $\epsilon_{2}$ must be much smaller than unity. On the other hand, if the accretion system has
much less angular momentum at larger radii or the accretion disc is highly viscous at larger radii, the flow 
may have even less angular momentum at inner radii. In such cases if the fluid moves radially without change in angular momentum, the 
angular momentum distribution is represented by $q=2$. Thus the flow with $q$ more close to $2$ will have less angular kinetic 
energy. Because of this the radial drift velocity will be larger than the the corresponding rotational velocity, thus $\epsilon_2$
will be larger than unity. In this case, the gravitational energy will be much larger than the angular kinetic energy and
hence the magnitude of $\epsilon_1$ will be much larger than unity. Apart from the above reasoning, range
of $\epsilon_{1}$, $\epsilon_{2}$ and $j$ must be consistent with physically acceptable values of $T$, $B$ and $M$. 

\subsubsection{Numerical results}

\begin{figure}
\centering
\includegraphics[width=0.90\columnwidth]{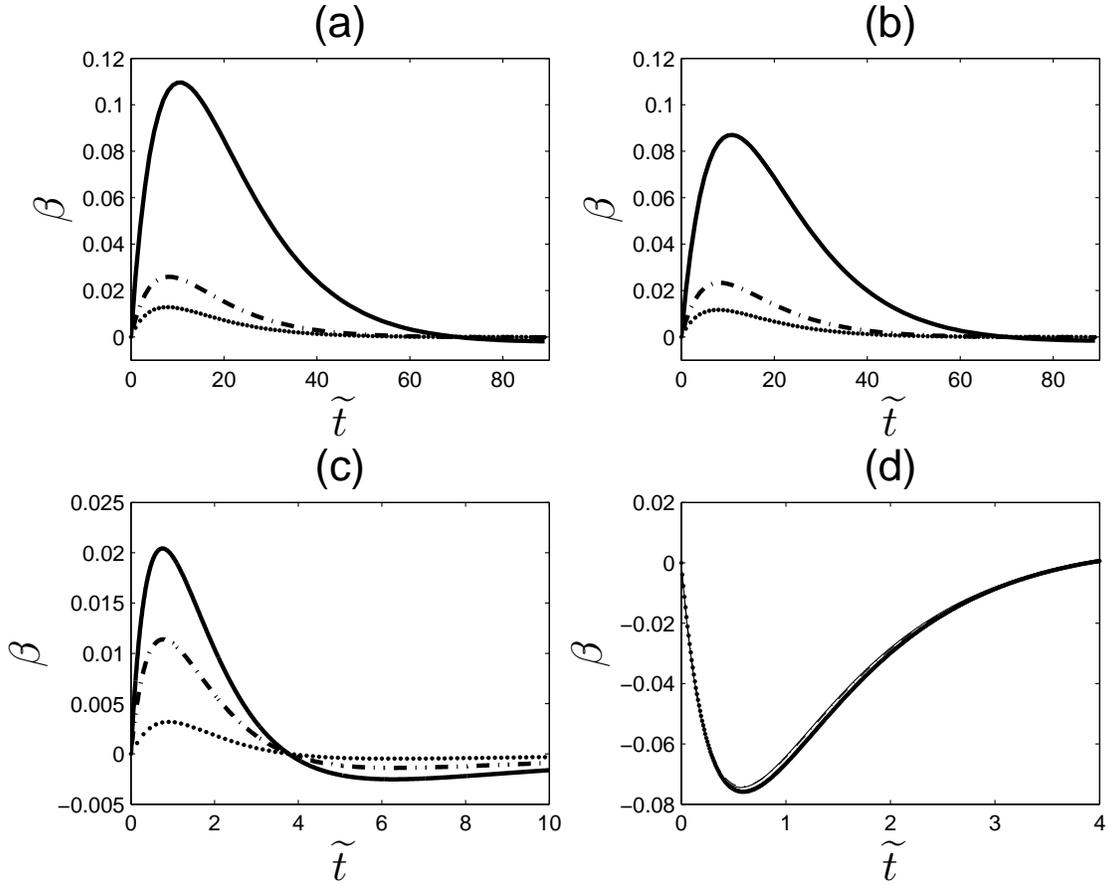}
\caption{The temporal evolution of the perturbation of the viscosity parameter is shown.
The solid, the dashed-dotted and the dotted curves in panels (a) and (b) correspond to $q=1.5$, $q=1.6$ and $q=1.7$, respectively,
whereas they correspond to $q=1.7$, $q=1.8$ and $q=1.9$, respectively in panels (c) and (d). We set $\alpha \sim 10^{-2}$ for (a) and (c),
whereas $\alpha \sim 0.99$ for (b) and (d). For panels (a) and (b) $\epsilon_{1}=-1.1$, $\epsilon_{2}=0.01$, and
for (c) and (d) $\epsilon_{1}=-3$, $\epsilon_{2}=1.125$. We set $j=0.9$ and $v(t=0)=0.5$, throughout. The horizontal axes in all the panels denote the time measured
in units of the rotational time-scale (see Eq.~(\ref{t-scale})).}
\label{neul1}
\end{figure}

By taking into account all the points presented previously, we numerically solve the dynamical set of equations, given
in Eqns.~(\ref{sig})-(\ref{bet}). Here we only show the temporal variations of the quantity $\beta$ and plot, in Fig.~(\ref{neul1}),
its time-evolution for variety of parameters. The four different panels in Fig.~(\ref{neul1}) correspond to four classes
of parameters, as described below:

\begin{enumerate}
\item [(i)] \emph{Keplerain/Nearly Keplerian, low viscosity case (Fig.~(\ref{neul1}a)) :} The solid, the dashed-dotted and the dotted curves in
Fig.~(\ref{neul1}a) correspond to $q=1.5$, $q=1.6$ and $q=1.7$, respectively. For all the three curves, we chose $\alpha \sim 10^{-2}$,
$\epsilon_1=-1.1$, $\epsilon_2=0.01$ and $j=0.9$.
\item [(ii)] \emph{Keplerain/Nearly Keplerian, high viscosity case (Fig.~(\ref{neul1}b)) :} The solid, the dashed-dotted and the dotted curves in
Fig.~(\ref{neul1}b) correspond to $q=1.5$, $q=1.6$ and $q=1.7$, respectively. For all the three curves, we chose $\alpha \sim 0.99$,
$\epsilon_1=-1.1$, $\epsilon_2=0.01$ and $j=0.9$.
\item [(iii)] \emph{Sub-Keplerian, low viscosity case (Fig.~(\ref{neul1}c)) :} The solid, the dashed-dotted and the dotted curves in
Fig.~(\ref{neul1}c) correspond to $q=1.7$, $q=1.8$ and $q=1.9$, respectively. For all the three curves, we chose $\alpha \sim 10^{-2}$,
$\epsilon_1=-3$, $\epsilon_2=1.125$ and $j=0.9$.
\item [(iv)] \emph{Sub-Keplerian, high viscosity case (Fig.~(\ref{neul1}d)) :} The solid, the dashed-dotted and the dotted curves in
Fig.~(\ref{neul1}d) correspond to $q=1.7$, $q=1.8$ and $q=1.9$, respectively. For all the three curves, we chose $\alpha \sim 0.99$,
$\epsilon_1=-3$, $\epsilon_2=1.125$ and $j=0.9$.
\end{enumerate}

\noindent
The initial value of the radial velocity perturbation was identical for all the
cases described above, which was chosen to be $0.5$.
Thus we have all the four regimes of parameters from high angular momentum, low viscosity
flow to low angular momentum, high viscosity flow. In all these regimes, we notice that
the flow with lower $q$ value has a stronger response to the same initial trigger,
with all other parameters being the same. This is because for the flow with lower $q$, the angular
momentum of the base flow is larger and to nullify the effect of an external trigger,
relatively more angular momentum has to be dissipated, and hence the viscous stress
has to increase by a larger amount, as characterized by $\beta$. 

Generally, radiatively inefficient sub-Keplerain components of accretion discs
are thought to be highly viscous, and in such situations the huge radial velocity
advects the fluid much before the radiative mechanisms could cool the system.
In Fig.~(\ref{neul1}d), the extreme case of flow with low angular momentum and
high viscosity is shown. Such a disc has already dissipated much of its angular
momentum and is rapidly heading towards an inherent unstable situation. The only way
to stabilise the system is by channelising the available external energy input to
reduce the viscous dissipation and radial velocity. Thus the fluctuation in viscosity, 
$\beta$ shows a decreasing tendency in this extreme case.

We note that there are two time scales involved in the problem, one, the radial drift time
of the particle, denoted by $\tau_D$, and the other, the saturation time of the perturbation,
denoted by $\tau_S$. In our scaling convention, $\tau_D=1/\sqrt{\epsilon_2}$ and
$\tau_S$ is defined as the time that it takes for the perturbation ($\beta$) to reach its
extremum value. The mechanism proposed here to produce a variation in the viscosity
parameter will be appreciable only if the saturation time is smaller than or
of the order of radial drift time. This is because if the
saturation time is much larger than radial drift time the fluid particle will
be drifted to the inner orbit much before the perturbation 
grows appreciably and hence the mechanism is inefficient to have any practical
relevance. We see from the above mentioned cases that under proper choice of
parameters, the condition $\tau_S \lesssim \tau_D$ is always met. Thus we conclude that this 
mechanism is physically efficient to produce an appreciable variability in viscosity.

\subsection{Power Law Growth}

A simple analytical model for the statistical theory (random saturation) of a nonlinear growth process was first introduced in
astrophysics by \cite{Fer49} in the context of cosmic rays. The most common growth processes in nature are either exponential 
growth or power law growth. Processes where a discrete number of basic units interact multiplicatively to form an 
avalanche (such as nuclear chain reactions) are well represented by an exponential growth law. A power law growth commonly occurs when 
the process scales with expansion in spatial domain (a brilliant account of such models can be found in \cite{Asc11}). In our case, the power emitted 
by the system is due to the process of viscous dissipation. If we neglect the density fluctuation, which is
always very small compared to the mean density, the power emitted (after subtracting the background power) by the process is 

\beq
 P\left(t\right) \,=\, \left( \Sigma T r \frac{\partial \Omega }{\partial r} \right) \beta \left(t\right)
\eeq

\noindent
The power emitted by the system is directly proportional to $\beta$, where in scaled
units at $\widetilde{r} \, = \, 1$, the proportionality constant depends on radial location through $q, j, \epsilon_1 \, \rm{and} \, \epsilon_2 \,$.

\begin{figure}
\includegraphics[width=0.90\columnwidth]{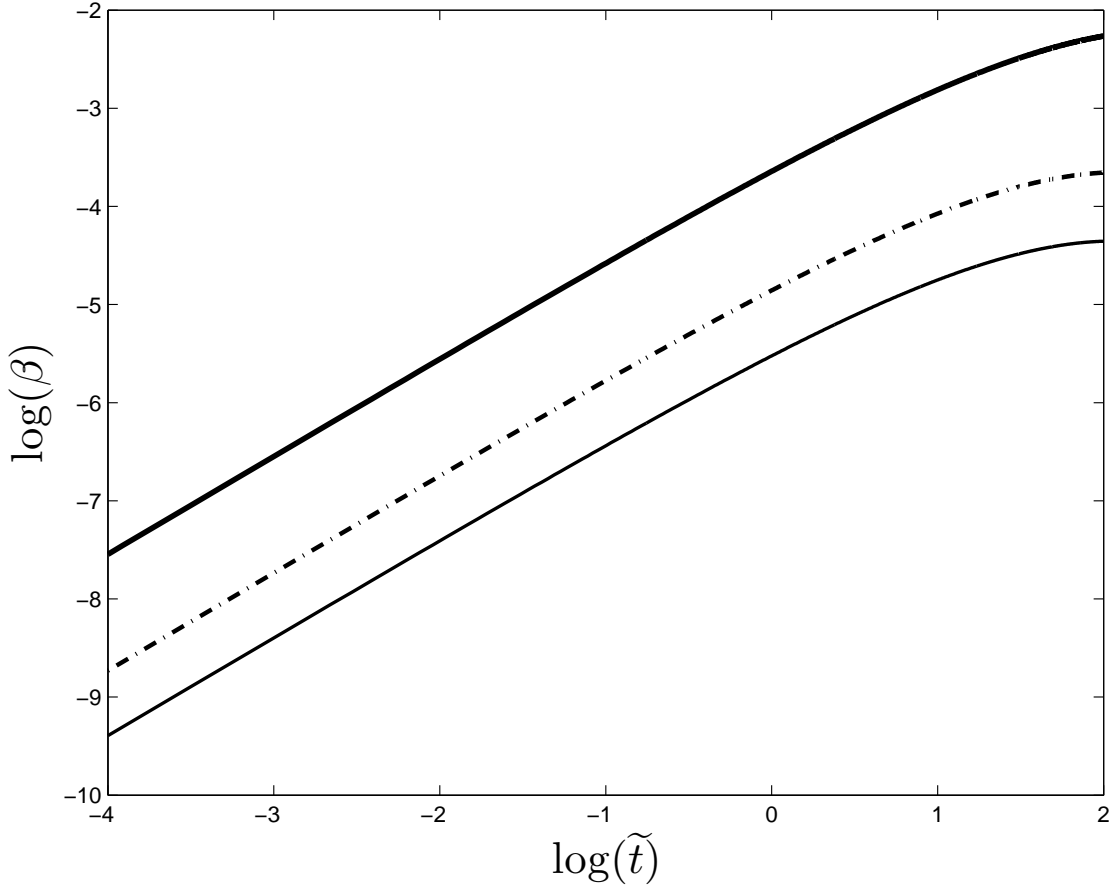}
\caption{Plot for the growing phase of $\beta$ for the Keplerian base flow ($q=1.5$) with 
$\alpha \sim 10^{-2}$, $j=0.9$ and $\epsilon_1=-1.1$ with $\epsilon_2\,=\,10^{-2}$ (thick line), $10^{-3}$ 
(dot dashed line), $10^{-5}$ (thin line).}
\label{log-log}
\end{figure}

As noted earlier the time scale of growth of the viscosity fluctuation is always less than or of the order of  
the radial drift time (advection time). Hence if we model the viscosity fluctuation as a nonlinear `process' 
of extracting energy, the relevant time of the process is the saturation time $\tau _S$, defined as the 
time when $\beta$ is extremum. In Fig.~(\ref{log-log}), $\log{(\beta)}$ as a function of
$\log{(\widetilde{t})}$ has been plotted upto the saturation time, $\tau_S$.
The curve is a straight line until it approaches the saturation time $\tau_S$.
From now onwards we represent $\widetilde{t}$ by $t$ for simplicity.
Hence for the time interval $0 \leq t \leq \tau_S$ we can approximately write 

\beq
\beta \,=\, \beta _{0} \,  t^{\gamma} 
\eeq

\noindent
where $\gamma$ is a positive number. The value of $\beta_0$ and $\gamma$ can be found by
fitting the curve numerically. The 
analytical expression given above deviates from the curves shown in Fig.~(\ref{log-log})
only when $t$ approaches the saturation time. This is because the 
power law growth cannot capture the saturation of the perturbation. Nevertheless the power law is the best analytical 
choice and the physics is well represented until $t \leq \tau_S$. The power law nature of the growing phase of the process is an indicator of its 
strong dependence on spatial location. Our perturbation analysis is done at a particular radial location of the general 
flow domain of an accretion disc, and the strong dependence of the perturbation on radial location is evident 
through the restrictive range of values for the free parameters. Hence

\begin{eqnarray}
P\left(t\right) &\,=\,& P_{0} \, t^{\gamma} \qquad \mbox{for}\qquad
(0 \leq t \leq \tau_S) \nonumber\\[2ex]
&& \mbox{where} \qquad P_0 \, = \, q \left( \frac{\epsilon_1}{j-2} \, + \, \frac{1}{j-2q+1} \, - \, \frac{j \epsilon_2}{j+1} \right) \, \beta_0
\end{eqnarray}

\noindent
Therefore the total energy emitted by the system in the time interval $0 < t < \tau$ is
\beq
 E\left(\tau\right) \,=\, \int_{0} ^{\tau} P \left( t \right) \, dt
\, = \, P_{0} \int_{0} ^{\tau} t^{\gamma} \, dt
\, = \, \left(\frac{P_0}{\gamma \,+ \, 1} \right)  \, \, \tau ^{\gamma +1}
\eeq

\noindent
The real probabilistic quantity in this theory is the time $\tau$ up to which the `process' continues. The probability distribution for 
$\tau$ is beyond the scope of the standard hydrodynamic theory. If the dynamical equation for
the probability distribution is thought of as a 
diffusion equation in $\tau$ then the probability distribution has the form
(see e.g. \cite{Asc11}),

\begin{eqnarray}
N\left(\tau\right) \, d\tau\, \propto \, e^{-a \tau} \, d\tau 
\end{eqnarray}

\noindent
Thus the probability distribution becomes 

\begin{eqnarray}
N\left(E\right) \, dE \propto \, E^{- \left(\frac{\gamma}{\gamma \, + \, 1}\right)}  \, \, e ^{-n \, E^{\frac{1}{\gamma \, + \, 1}}}\, dE,  \, \,\mbox{where} \, \, n \, = \, a \, \left( \frac{\gamma +1}{P_0}\right)^{1/(\gamma +1)} 
\end{eqnarray}

\noindent
where $N$ is defined as the probability that the nonlinear process extracts an amount of energy $E$ from the 
background shear and $a \, \sim \,1/\tau_S $.

We note that there is a difference between the 
`nonlinear process' discussed in \cite{Asc11} and the one we analysed in this work.
In the analytical models of \cite{Asc11} it is 
assumed that the growing phase of the `process' could either be an exponential
or a power law function of time, and the process remains nonlinear up to the saturation time. 
Whereas, we numerically find that the
growing phase of our process is strictly a power law in time. In \cite{Asc11},
final dissipation is treated as a linear process, whereas, in our real physical
system another time scale enters which is the 
radial drift time, $\tau _D$. It would be physically irrelevant to consider any growth
of the perturbation beyond $\tau_D$, as the fluid particle itself would have moved to inner
orbits due to radial advection at times of order $\tau_D$.
Since $\tau _S$ is always comparable to or less than
$\tau_D$, we need to focus only 
on the growing phase. The real probabilistic quantity is $\tau$, which signifies
how long the `process' continues in the growth phase.
It is this probability which we assumed to be as an exponential function.
We also note that the saturation time of the growth process studied in this work
shows strong dependence on the radial location through the parameters
$q, j, \epsilon_1 \, \rm{and} \, \epsilon_2$.

\section{Conclusion}

In the above analysis we have proposed a mechanism which causes fluctuations in the $\alpha$
viscosity parameter for a general angular momentum distribution, Keplerian or non-Keplerian
($1.5\leqq q < 2$). We began by considering a vertically integrated, axisymmetric, steady-state
accretion flow (base flow), in which, we focussed on an annular region of the accretion disc.
The global flow domain can be thought of as a collection of such annular regions.
We modelled the steady state flow variables in the annular region as the power law
in the radial coordinate. We assumed the closure model proposed by \cite{SS73} and
treated the viscosity parameter $\alpha$ as a continuum variable. Our aim was to study
how the background $\alpha$ could change in response to any perturbation on the
radial velocity and density fields. 
Ignoring any instantaneous perturbation of the base angular momentum, we perturbed
the density and radial velocity fields, which enabled us to obtain an expression for the
evolution of the perturbation as a function of the viscosity parameter. We reduce the perturbation
equations to a set of nonlinear dynamical systems of equations, by specializing to the
case when the Lagrangian derivatives (defined with respect to the radial velocity field)
of the perturbations in the flow variables vanish. This special choice enabled us to
reduce the coupled nonlinear perturbation equations to a set of first-order, coupled,
nonlinear, autonomous, dynamical equations in time. It may be worthwhile to remark that
such analysis would provide enough motivation to isolate this class of perturbations in
future numerical experiments whose effects are interesting as demonstrated in the present
work. We found that the base $\alpha$
could change due to its nonlinear dynamical coupling to other fluid variables and perturbations
of them. Thus we treated the perturbation of $\alpha$ (denoted by $\beta$) as a passive/slave
variable in the sense of dynamical system theory.

Such studies have direct implications on the observed variability in X-ray luminosities
from accretion discs. \cite{Lyu97} has shown that local fluctuations of the $\alpha$
parameter could cause the mass accretion rate at inner radii to vary and hence
cause a temporal variation in energy dissipation. We have shown in the present
work that the fluctuations in $\alpha$ could come about due to perturbations
on the mass accretion rate at some radius, and they could behave in a complicated
manner as a function of radial distance from the central object. Our main conclusions
may be stated as follows:

\begin{enumerate}
 \item [(i)] We demonstrate that the viscosity parameter in the accretion disc
 can change in appreciably short and astrophysically relevant time scale.
 \item [(ii)] We find that the saturation time ($\tau_S$) of the perturbation is
 smaller or of the order of the radial advection time scale ($\tau_D$). This is an
 important point because if the saturation time is larger than radial drift time,
 the fluid particle will be drifted to the inner orbit much before the perturbation grows
 appreciably, and hence the mechanism would be inefficient to have any practical relevance.
 Thus we conclude that this mechanism proposed here is physically efficient to
 produce appreciable variability in viscosity.
 \item [(iii)] The perturbation ($\beta$) shows growth, saturation and eventual degradation,
 which is a non-trivial result due to nonlinear nature of the system of equations. This offers
 a complete physical picture of nonlinear evolution of the perturbation.
 \item [(iv)] We have demonstrated that the variability in $\alpha$ in the growing phase
 is an exact power law in time.
 \item [(v)] We have also demonstrated that not only the time scale of the fluctuation, but
 also the amplitude of the fluctuation strongly depends on radial location. The time scale
 of the fluctuation is found to be independent of the initial trigger.
\end{enumerate}

A stable accretion system tries to maintain the steady mass flow across all radii. Any cause, internal or external which 
disturbs this steady state, gets quickly nullified by sufficient variation of viscous dissipation. We can think of the local fluctuation in $\alpha$ and the associated change in mass
accretion rate as a chain of coupled processes across radii $r_1 > r_2 > ... r_i$.
A zero initial trigger in active modes (density perturbation and radial velocity perturbation)
could cause no change in $\alpha$. Let us consider an annular region at radius $r_1$ where
the mass accretion rate is slightly changed/perturbed due to some process happening at
outer radii, say, due to the fluctuation in $\alpha$ proposed by \cite{Lyu97}. This will
seed the initial trigger in radial velocity perturbation at radius $r_1$. Thus the system
will evolve according to the set of equations described in \textsection~(2.3). This induces
local fluctuation in $\alpha$ at $r_1$ which would cause change in the mass accretion rate
in the accretion time scale at $r_1$, $\tau_D \left( r_1 \right)$. This will 
excite a fluctuation in $\alpha$ at the next inner orbit at $r_2$ 
in a time scale $\tau_D\left( r_2 \right)$. 
This will cause a larger change in mass accretion rate compared to the change in mass
accretion rate at $r_1$ (this is because $\beta$ is independent of local scaling and
the real time involved in the fluctuation decreases as radius decreases). This chain of
process continues until the radius $r_i << r_1$, where most of the energy is released. The
variation in the mass accretion rate at the inner orbit is therefore a cumulative effect
of $\alpha$ fluctuations at different outer radii and hence a small variation 
of $\alpha$ at the outer radius could cause a huge effect at the inner orbit. The amplitude
of the fluctuation in $\alpha$ depends on the radial location through $j, q, \epsilon_1$ and
$\epsilon_2$. But these parameters at each radius should be found from a global 
solution of conservation laws. Thus each accreting system found observationally, should
be explained by the branch of a global solution which can account for the observed time variability in luminosity (Appendix~A).

\section*{Acknowledgments}

We are grateful to K. Subramanian, D. Bhattacharya and R. Misra of IUCAA for
many useful suggestions and discussions. We also thank the referee for useful
comments which helped improve the quality of the manuscript. SRR gratefully acknowledges
the visiting associateship and hospitality provided by IUCAA.

\section*{Appendix A}

The accreting matter is a mixture of ions and electrons. The mass content of the fluid is mainly due to ions and the radiative process
 (cooling mechanism) is effectively due to electrons. In the most general case the ion and electron temperatures are different. Let 
$T_i$ be the ion temperature and $T_e$ be the electron temperature and $T$  be the
effective temperature. Then the equation of state is \citep{NY95,RM10}

\beq
P \, = \, \Sigma \, T \, = \, \frac{\Sigma}{f c^2} \left( \frac{K_B T_i}{\mu _i m_i} \, + \, \frac{K_B T_e}{\mu _e m_e} \right)
\label{Papp}
\eeq

\noindent
($T$ in the units of light speed squared and $T_i$, $T_e$ measured in the units of Kelvin. $\mu_i$, $\mu_e$ are the effective 
molecular weights of ion and electron. $m_i$ is the mass of proton and $m_e$ is the mass of electron. $f$ is the ratio of gas 
pressure to the total pressure. $K_B$ is the Boltzmann's constant and $c$ is the speed of light in vacuum.)

For a general advective flow the turbulent energy dissipated ($Q_{vis}$) is partly advected radially ($Q_{adv}$) and partly 
radiated away ($Q_{rad}$) by electrons. $Q_{vis} \, = \, r \frac{\partial \Omega}{\partial r} W^{\phi r}$ and 
$Q_{adv} \, = \, Q_{adv} \left( \Sigma, V, T\right)$. $Q_{rad}$ depends on the radiative process we choose. In the case of a sub-Keplerian 
disc $Q_{rad}$ is mainly bremsstrahlung, synchrotron processes and inverse
Comptonization due to soft synchrotron photons \citep{NY95,RM10}. In the case of a Keplerian disc $Q_{adv}$ is neglected and it is assumed that the 
local energy dissipated is radiated thermally. In both cases $Q_{rad} \, = \, Q_{rad} \left( \Sigma, T_e \right)$. 
Thus in general we can write

\beq
Q_{vis} \left( \alpha, \Sigma, \Omega, T \right) \, = \, Q_{adv} \left( \Sigma, V, T \right) \, + \, Q_{rad} \left( \Sigma, T_e \right)
\label{Qvis}
\eeq

\noindent
Solving the above equations (\ref{Papp}) and (\ref{Qvis}) simultaneously we can find the ion temperature ($T_i$), electron temperature ($T_e$) and the corresponding 
radiation output. In the case of perturbed system discussed in the main text the fluctuation in radiation output due to $\alpha$ fluctuations 
can be computed.  

\section*{Appendix B}

\begin{eqnarray}
F_1 &=& \frac{B_2 f_1 \, - \, B_1 f_2}{A_1 B_2 \, - \, B_1 A_2} \nonumber \\[2ex]
F_2 &=& \frac{A_1 f_2 \, - \, A_2 f_1}{A_1 B_2 \, - \, B_1 A_2} \nonumber \\[2ex]
F_3 &=& \frac{B_1 A_3 f_2 \, - \, A_3 B_2 f_1 \,+\, (A_1 B_2 \, - \, A_2 B_1) f_3 }{A_1 B_2 C_3 \, - \, B_1 A_2 C_3}
\end{eqnarray}

\begin{eqnarray}
A_1 &=& \widetilde{v} \;;\qquad
B_1 \,=\, (1 + \widetilde{\sigma}) \;;\qquad
C_1 \,=\, 0 \nonumber \\[2ex]
A_2 &=& \widetilde{T} \;;\qquad
B_2 \,=\, \epsilon_2 (1 + \widetilde{\sigma}) \, \widetilde{v} \;;\qquad
C_2 \,=\, 0 \nonumber \\[2ex]
A_3 &=& \widetilde{T} (\alpha + \beta) \;;\qquad
B_3 \,=\, 0 \;;\qquad 
C_3 \,=\, \widetilde{T} (1 + \widetilde{\sigma}) 
\end{eqnarray}

\begin{eqnarray}
f_1 &=& \epsilon_2^{1/2} \left[ (1 - j) \widetilde{\sigma} \,+\,
j \widetilde{v} \,+\, \widetilde{\sigma} \widetilde{v} \right] \nonumber \\[2ex]
f_2 &=& -\epsilon_2^{3/2} j\, \widetilde{\sigma} \,-\, \epsilon_2^{1/2} \widetilde{\sigma} \,-\,
\epsilon_2^{3/2} j \widetilde{v} (1 + \widetilde{\sigma}) \,+\, \epsilon_2^{1/2} \widetilde{\sigma}
\left[\frac{2 j^2\,\epsilon_2}{j+1} \,+\, \frac{2(1-q)}{j-2q+1} \,-\, \frac{\epsilon_1}{j-2}\right]
\,-\, \epsilon_2^{1/2} \epsilon_1 \widetilde{\sigma} \nonumber \\[2ex]
f_3 &=& -(3-q-j) \epsilon_2 \widetilde{\sigma} \,+\, e \widetilde{\sigma} \,-\, 
\frac{j(1-j)}{1+j} \epsilon^{3/2} \beta \,+\, \frac{(j-2q+3)}{(j-2q+1)} \epsilon_2^{1/2} \beta \,+\,
\frac{j}{j-2}\epsilon_1 \epsilon_2^{1/2} \beta \nonumber \\[2ex]
&& -\,\frac{2j(1-j)}{1+j} \epsilon_2^{3/2} \beta \widetilde{\sigma} \,+\,
\frac{2(2-q)}{(j-2q+1)} \epsilon_2^{1/2} \beta \widetilde{\sigma} \,+\,
\frac{\epsilon_1 \epsilon_2^{1/2}}{j-2} \beta \widetilde{\sigma} \,+\,
(2-q)\epsilon_2 \left[\widetilde{\sigma} +  (1 + \widetilde{\sigma})\widetilde{v}\right]
\end{eqnarray}

\end{document}